\documentclass[a4paper,
               keeplastbox,   % flushend option: not to un-indent last line in References
               ]{jacow}
\usepackage{pdfpages,multirow,ragged2e} %
\makeatletter%
	\ifboolexpr{bool{xetex}}
	 {\renewcommand{\Gin@extensions}{.pdf,%
	                    .png,.jpg,.bmp,.pict,.tif,.psd,.mac,.sga,.tga,.gif,%
	                    .eps,.ps,%
	                    }}{}
\makeatother

\ifboolexpr{bool{xetex} or bool{luatex}} % test for XeTeX/LuaTeX
 {}                                      % input encoding is utf8 by default
 {\usepackage[utf8]{inputenc}}           % switch to utf8

\usepackage[USenglish]{babel}
\usepackage{blindtext}

\usepackage{siunitx}
\ifboolexpr{bool{jacowbiblatex}}%
 {%
  \addbibresource{refs.bib}
 }{}
\begin{document}

\title{Recent developments in LLRF and its controls at CERN Linac4}
%\title{Development of LLRF for CERN’s Linear Accelerator 4}

%%%  ArXiv
%%%  https://arxiv.org/html/1909.06754 
%%%  LLRF2019/166

\author{B.~Bielawski\thanks{bartosz.bielawski@cern.ch}, P.~Baudrenghien, R.~Borner, CERN, Beams Department, Geneva, Switzerland \\}
\maketitle

\begin{abstract}
At the end of the Large Hadron Collider’s Run 2, CERN’s Proton Injector Linac 2, commissioned in 1978, delivered its final beam in December 2018. For Run 3, from March 2021, a new H$^{-}$ will take over the role: Linac4. The machine has been producing test beams since 2013 and in 2016 it reached its design 160 MeV energy with 20 mA beam current. Since then several improvements have been made which enhance the LLRF performance and stability.
In this paper the structure of the machine and the control system are presented. Problems arising from different RF station types are described and our solutions explained, along with new features recently added to the LLRF. An Adaptive Feed Forward implemented as a flexible hybrid hardware-software solution is described, and first results of this application running on a laboratory test stand are presented. Addition of the set point modulation needed for longitudinal phase-space painting is discussed.
Finally, software tools for automation of setting up, monitoring and operations are described. As the hardware and low-level software are still reaching maturity, these are only briefly introduced here.
\end{abstract}

\section{Introduction}
Linac4, the newest of CERN's linear accelerators, is being prepared to take over the role of the initial injector for the proton accelerator chain. The most important parameters of the new machine can be found in Table \ref{tab:linac4_parameters} \cite{ln4-tdr}.

\begin{table}[h]
 \caption{Linac4 --- main parameters.}
  \centering
  \begin{tabular}{ll}
    \toprule
    \textbf{Parameter} & \textbf{Value} \\ 
    \midrule
    length & 80 \si{m} \\
    accelerated ion & H$^-$\\
    RF frequency & 352.2 \si{MHz}\\
    output energy & 160 \si{MeV}\\
    avg. beam current & 40 \si{mA} \\
    pulse length & 600 \si{us} \\
    repetition rate & 0.8$\overline{3}$ \si{Hz} \\
    \bottomrule
  \end{tabular}
  \label{tab:linac4_parameters}
\end{table}

From the radio frequency and control system point of view Linac4 is not a homogeneous accelerator. One can distinguish at least five different types of RF structures arranged in four groups:

\begin{itemize}
\item \textbf{L group} --- Low-Energy part (RFQ and 3 Pill-Box type bunching cavities),
\item \textbf{D group} --- Drift Tube Linac part (3 DTL cavities),
\item \textbf{C group} --- Cell-coupled Drift Tube Linac (7 CCDTL cavities),
\item \textbf{P group} --- Pi-Mode Structure cavities (12 cavities + 1 debuncher cavity).
\end{itemize}

Even within these groups not all cavities are the same or have the same drive system. 

During runs in 2018 and 2019 the linac has been delivering beams for many weeks. This time was used to study characteristics, test stability and identify possible improvements of the machine. 

\subsection{Low-level RF hardware modules}

Each of the RF amplifiers (that is all the RF related equipment together) is controlled by a single VME crate containing a Front-end Computer (FEC) and three main custom-designed LLRF modules \cite{kalman}: \textit{Cavity Loop}, \textit{Tuner Loop} and \textit{Switch and Limit}. All the modules contain FPGAs (Field-programmable Gate Array) and sets of analogue-to-digital (ADC) and digital-to-analogue converters (DAC). The presence of FPGAs allows in-circuit reconfiguration of hardware and continuous development of the module.

The \textit{Cavity Loop} regulates voltages from up to three antennas in the cavity and drives the cavity by a chain of amplifiers. Many recent improvements have been implemented in this card. 

The \textit{Tuner Loop} is responsible for keeping cavities on tune. By comparing the phase of the drive and antenna signals the module provides an error signal. Software reads out this error signal and issues commands to mechanical tuners (or water cooling for the RFQ) using a special Ethernet-enabled Motor Controller Interface (MCI).

The \textit{Switch and Limit} module has two main roles: the first one is to prevent the klystrons entering the saturation region. The second function is cutting the drive signal to the power amplifiers in case of an interlock.

Each of the crates is also equipped with a Crate Manager which monitors the hardware and relays trigger signals from the front panel to the backplane, and a Clock Distributor. The Clock Distributor is responsible for receiving the clock signal from the main reference line and generating harmonically related clocks for all I/Q demodulators and modulators (88.05 MHz [$\frac{f_{RF}}{4}$] for clocking ADC, DAC and FPGA, 330.187 MHz [$\frac{15}{16} \cdot f_{RF}$]  for local oscillator).

\subsection{Software}
\begin{figure}[!htb]
   \centering
   \includegraphics*[width=.7\columnwidth]{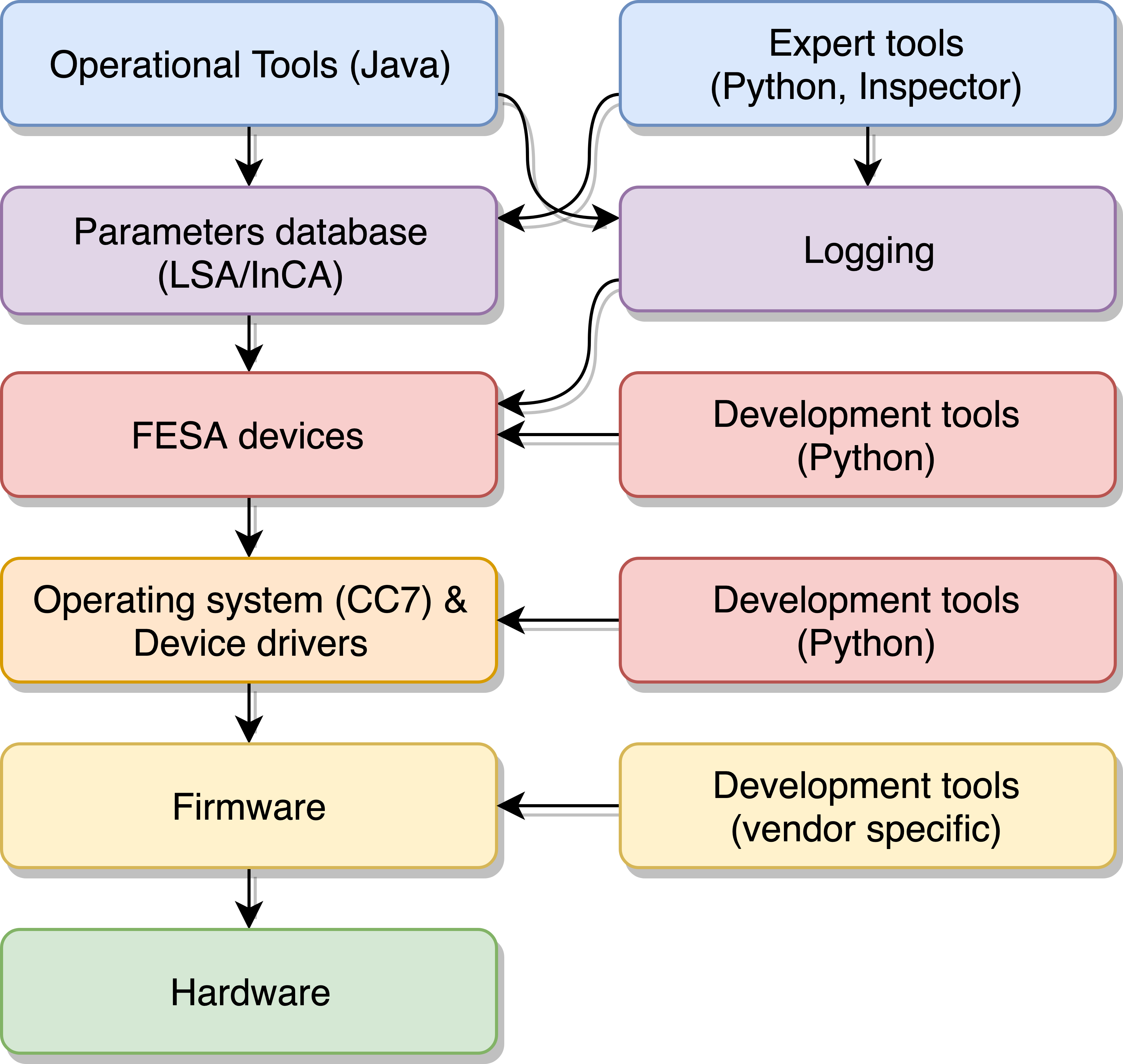}
   \caption{Typical software stack used at CERN.}
   \label{fig:software-stack}
\end{figure}

Multiple levels of software are present in large installations such as Linac4. A typical stack used at CERN is presented in Figure \ref{fig:software-stack}. The hardware, once fixed, is not changed very often. On the other hand, firmware can be updated quite often as new features are developed. Operating systems and Parameter Databases are provided by the CERN Controls and Operations groups. 

FESA devices are created using the Front-End Software Architecture (FESA) framework \cite{fesa} developed at CERN. The aim of this framework is to simplify the creation of software. Physical devices are usually represented by a set of corresponding software devices. 

During the commissioning process some extra tools are needed to quickly and reliably set up the system. In the case of Linac4 RF system some of these tools are written in Python which allows us to automate more things and speed up development.

\section{Improvements in Linac4 LLRF}

\subsection{Improved feedback controller}
%I'm lacking history input from Philippe at the moment
%Anirban Bhattacharyya - AFF (cavity model mentioned, no other trace),
%Jose Noirjean - posters, slides, paper with Philippe
%Javier Galindo - paper with Philippe
The specification for the maximum cavity voltage variation due to beam-induced voltage with 40 mA beam current is $\pm$1\% of the nominal voltage amplitude and $\pm$0.5\% in phase, above which beam quality is compromised. The machine was first commissioned in 2013 with a simple Proportional-Integral (PI) controller. A more robust solution, the Linear Quadratic Gaussian Regulator (LQG) was planned \cite{kalman} and has now been installed on all cavities, except for the debuncher.

The LQG, although much more complex, is capable of faster and more precise regulation. The Kalman predictor uses a model of the observed process to reliably estimate the next states of the system and this in turn means that the feedback is able to react faster. 

\begin{figure*}[!htb]
    \centering
    \includegraphics*[width=0.9\textwidth]{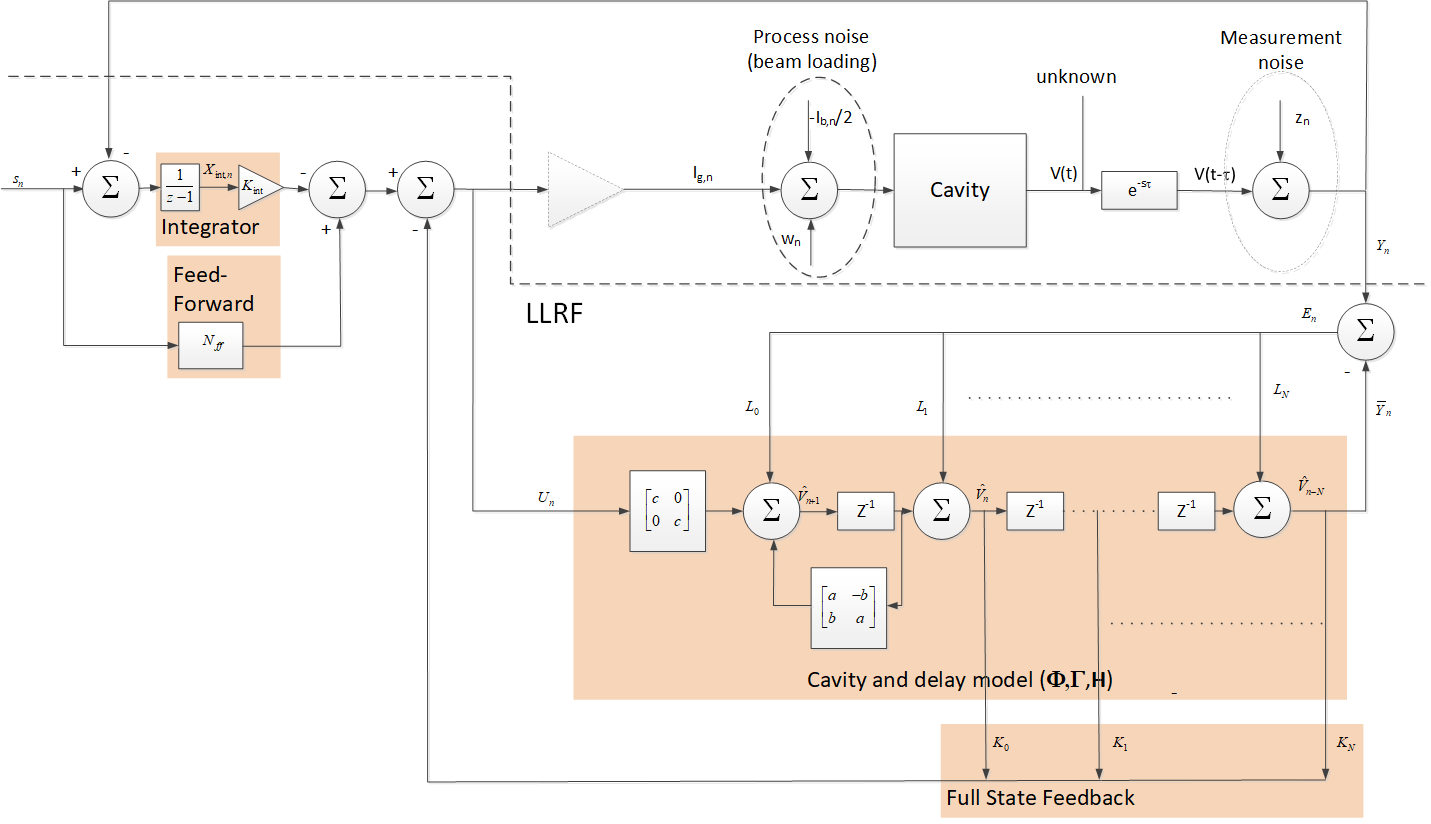}
    \caption{Internal structure of LQG/Kalman regulator.}
    \label{fig:lqr-kalman}
\end{figure*}

In this case the card has a model of the cavity (low-pass filter) and cables (delay line) built into the firmware. Since all the intermediate states are known in the estimator it is possible to use the state at the output of the cavity to achieve faster feedback. The internal structure is presented in Fig.~\ref{fig:lqr-kalman}.

The Cavity Loop card drives different types of cavities and therefore the model has to be tuned to match the type of cavity. During initial setup the RF team uses a set of scripts written in MATLAB and Python to automate the process. The script is aware of differences between RF lines with respect to filling times, loop delays and overall gains and can apply correct coefficients. 

In the near future it is planned to replace the remaining MATLAB code that was used to calculate Kalman gains with a pure Python implementation. Scanning of the parameter space is also considered for a future improvement.

\subsection{Adaptive Feed-Forward (AFF)}
Even with the improved LQG/Kalman feedback it is not possible to eliminate the transient beam-loading observed in the first 10 $\mu s$ at the head of the batch. Since this transient repeats at every pulse it is possible to apply a feed-forward solution to it \cite{kalman,fong-aff,aff}.
% This makes citations with multiple numbers in [] using commas or a dash [1,2] or [1-3]

The Linac4 beam consists of four batches one per Proton Synchrotron Booster (PSB) ring, with gaps in between. The AFF has therefore to correct four transients placed at the head of the four batches.

\begin{figure*}[!htb]
    \centering
    \includegraphics*[width=0.9\textwidth]{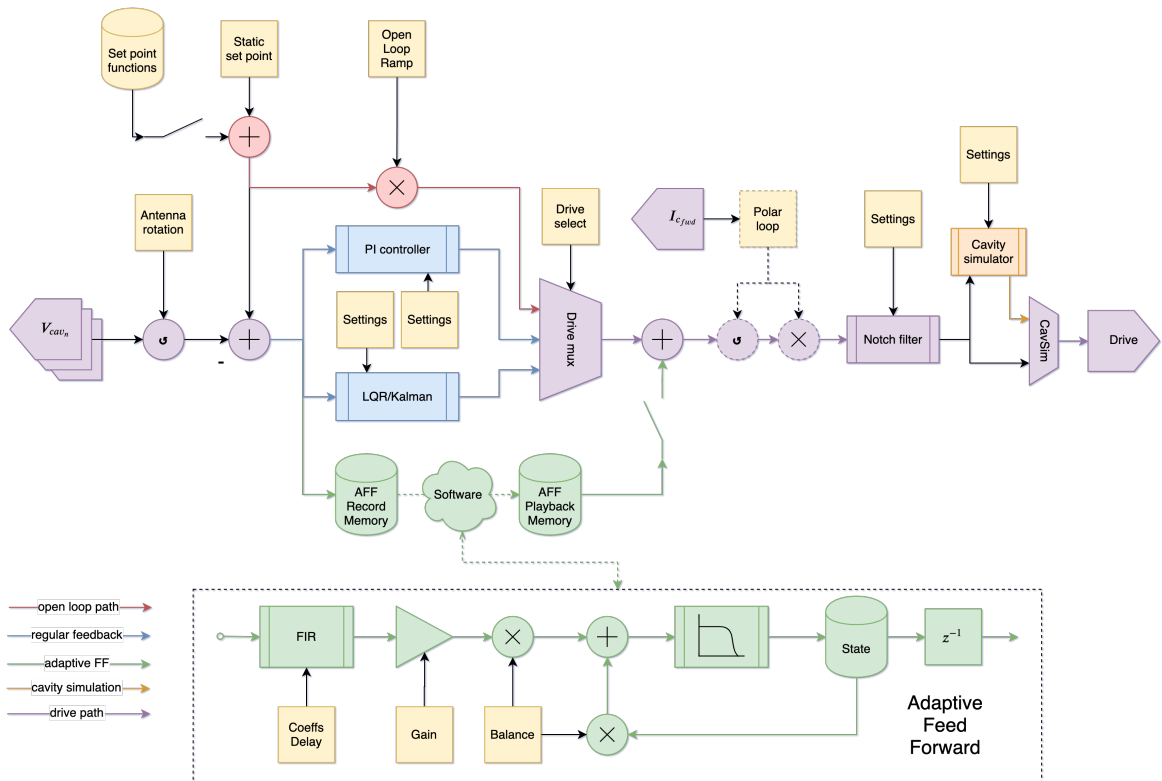}
    \caption{Block diagram of Linac4 Cavity Loops firmware \& software feedback and feed forward.}
    \label{fig:cavity-loops}
\end{figure*}

The hardware part is relatively simple and consists of four error and correction buffers, each 16 kS ($f_s = $ 22 \si{MHz}) long. Error buffers are read by the software at the end of a cycle, processed and saved as a state. At the beginning of the next cycle state buffers are loaded and values from them are added directly to the drive signal (see fig. \ref{fig:cavity-loops}, bottom, marked in green).

The state that is calculated and stored in the software is the correction that has to be applied to the drive signal. Every cycle the state is updated using the most recent measurements. The raw error signal is first filtered using an FIR filter (N $\leq$ 256), next a gain $G$ is applied:
$$e[n] = G \cdot \sum_{i=0}^{N-1} b_i \cdot x[n-i+L]$$
Constant $L$ represents the delay between the correction and the observation buffer expressed in processing clock cycles~($\frac{16}{f_{RF}}$). It allows the system to start applying a correction at the moment when the beam-loading starts but before it is observed. In a regular system such a filter would be non-causal, but in this case we are applying a correction to the next cycle, so the effective delay is slightly less than a full cycle.

% I have removed the temporal window from the equation because Philippe is not sure it's needed
The processed error signal is then combined with the previous state. The mixing ratio is specified by the balance $b$ parameter (controlling learning speed):
$$s_{n}^{(k)} = (1-b) \cdot s_{n}^{(k-1)} +  b \cdot e_{n}^{(k-1)}$$
The Adaptive Feed-Forward can be used with or without the feedback. Time series of the states vary in these two cases. Without the feedback the shape of the time series is rectangular which corresponds to the klystron power needed to maintain the voltage. When the feedback is on, the shape looks like an exponential decay curve.

\begin{figure}[!htb]
    \centering
    \includegraphics*[width=.88\columnwidth]{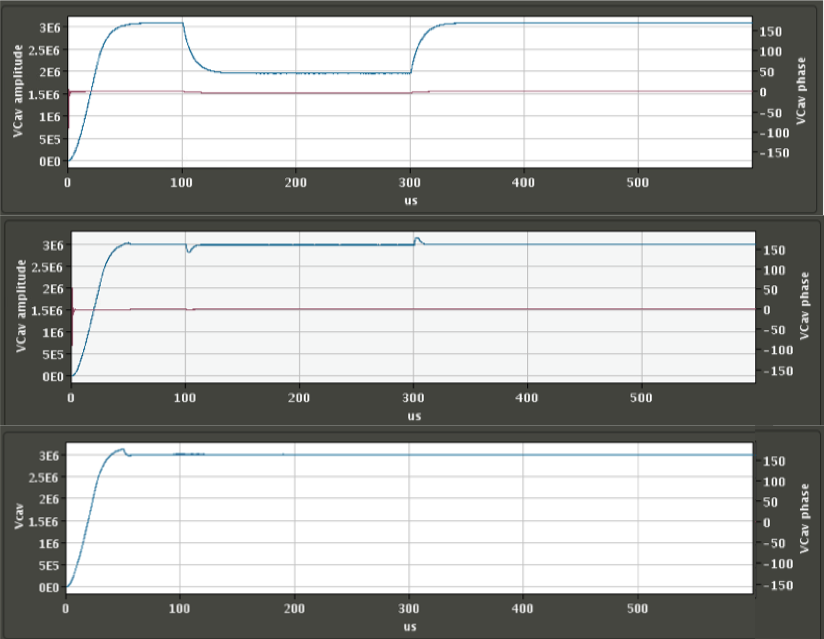}
    \caption{Cavity voltage in our laboratory test stand with simulated beam-loading: unregulated (top), with feedback (center), with feedback and AFF (bottom).}
    \label{fig:aff-waveforms}
\end{figure}

The overall result of applying AFF in our laboratory test stand can be seen in Fig.~\ref{fig:aff-waveforms}. The top waveform is the voltage in a cavity with a simulated beam-loading and no feedback. The middle waveform shows the voltage with beam-loading and feedback enabled. The bottom waveform has both feedback and AFF enabled.

For automatic setting up of the AFF a script in Python was written. The script applies excitation (via playback buffers) in the form of a step function, measures the delay of the system and calculates the FIR filter coefficients ${b_i}$. The AFF will be deployed and commissioned on all cavities in the coming weeks.

\subsection{Longitudinal painting}
Longitudinal painting is a technique which allows filling the longitudinal phase space of the PSB evenly, which in turn reduces multiple problems present with high-intensity beams \cite{psb-long-paint}. The implementation consists of modulation of the set-points for the last two PIMS (amplitude) and the debuncher (phase) (Fig. \ref{fig:cavity-loops}, top left).

In order to perform longitudinal painting the LLRF firmware and software had to be updated. During the 2019 run this newly developed feature will be tested in the machine.

\begin{figure}[!b]
    \centering
    \includegraphics*[width=.88\columnwidth]{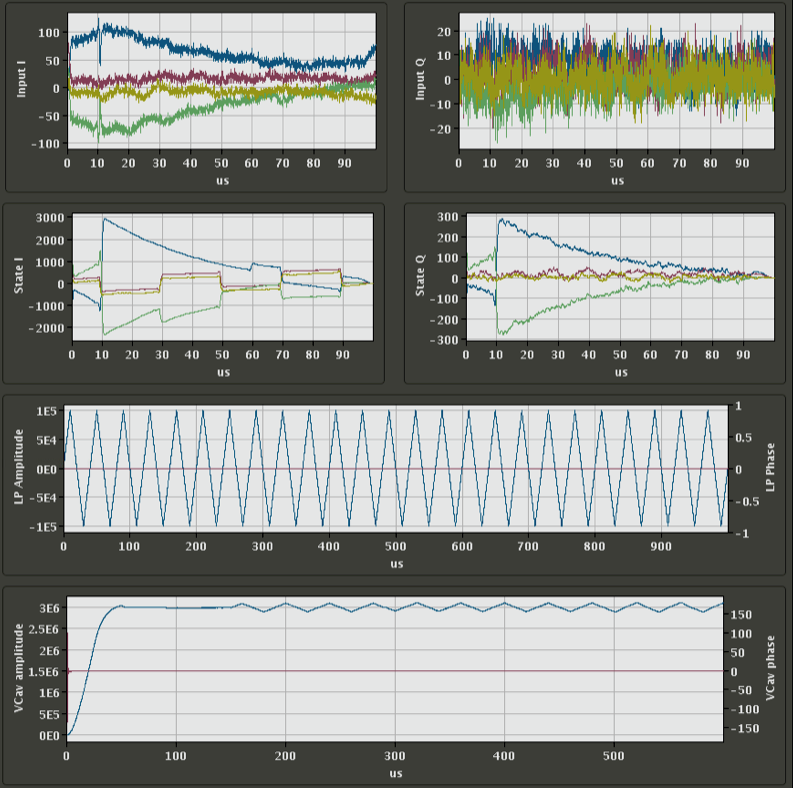}
    \caption{Voltage regulation with simulated beam-loading, AFF and longitudinal painting.}
    \label{fig:painting-and-aff}
\end{figure}

Figure \ref{fig:painting-and-aff} illustrates internal states of the Cavity Loop with AFF and longitudinal painting enabled. The first two top plots show the error signal as seen by the feedback (I and Q components). The second row shows the AFF state (drive correction, also I and Q). In the above two plots the four traces correspond to the four batches. It is worth noting the smoothness of the plot due to the applied filtering. The third row contains the function modulating the set-point and the last plot shows the voltage in the cavity. The shape of the modulation can be seen in the state buffer as the AFF corrects the drive signal for the sudden need of additional klystron power.

The 2019 run will be used to test the concept, hardware and software needed for longitudinal painting. Operational tools and shapes of wave forms will be determined when the first real measurements are available.

\subsection{High-level software}
The software that directly accesses the hardware is written in a compiled language such as C or C++. Higher level applications, operational and expert tools, are running on machines which have a lot of RAM and CPU power and can be written in Java or Python.

\begin{figure}[!t]
    \centering
    \includegraphics*[width=.8\columnwidth]{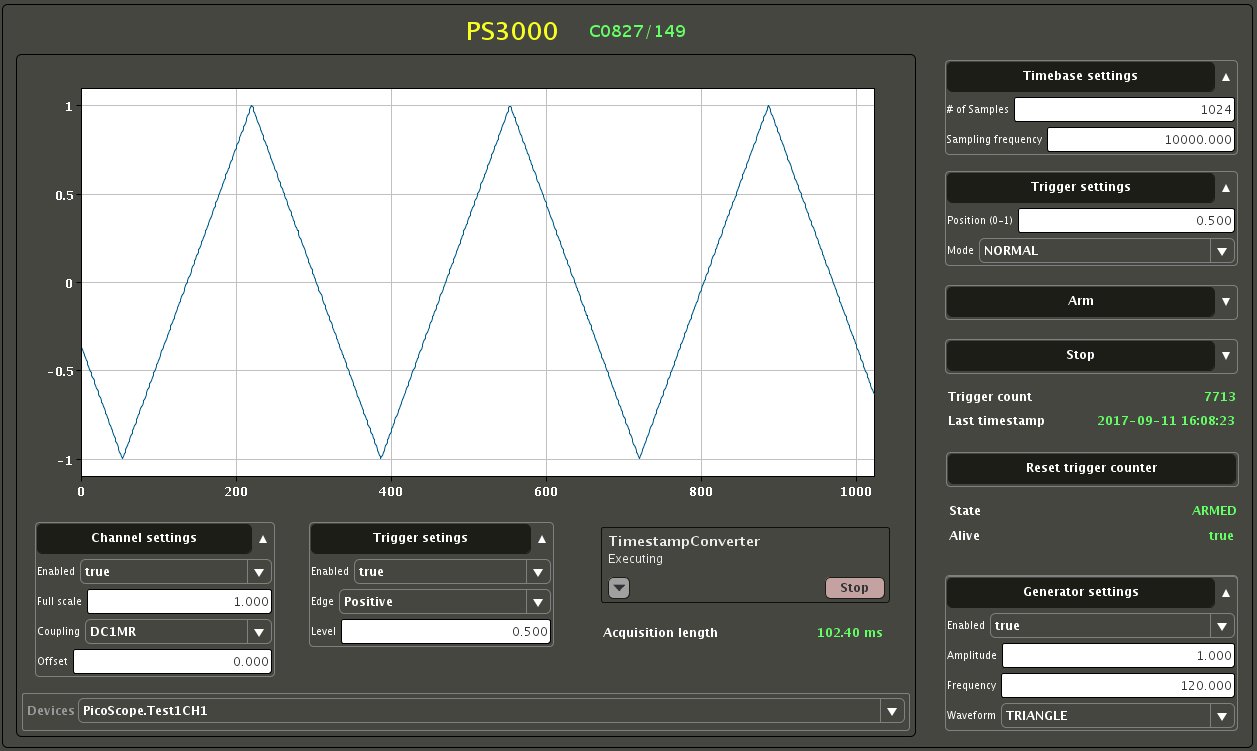}
    \caption{A generic scope interface using Inspector.}
    \label{fig:inspector-scope}
\end{figure}

One of the tools that is extensively used at CERN in the Radio Frequency group is Inspector \cite{inspector}. It is an environment that allows the creation of "panels" that are able to control and monitor properties exposed by FESA devices. Multiple different widgets are available such as dials, buttons, sliders, numeric inputs and plots. Figures \ref{fig:aff-waveforms}--\ref{fig:inspector-scope} show panels created using Inspector.

To increase flexibility it is possible to embed Python scripts in Inspector panels and call external programs. This feature was used to integrate another tool: the Linac4 RF Sequencer \cite{sequencer}. The sequencer is written in Python and its main purpose is to turn on and off the RF system in Linac4.

\begin{figure}[!htb]
    \centering
    \includegraphics*[width=.65\columnwidth]{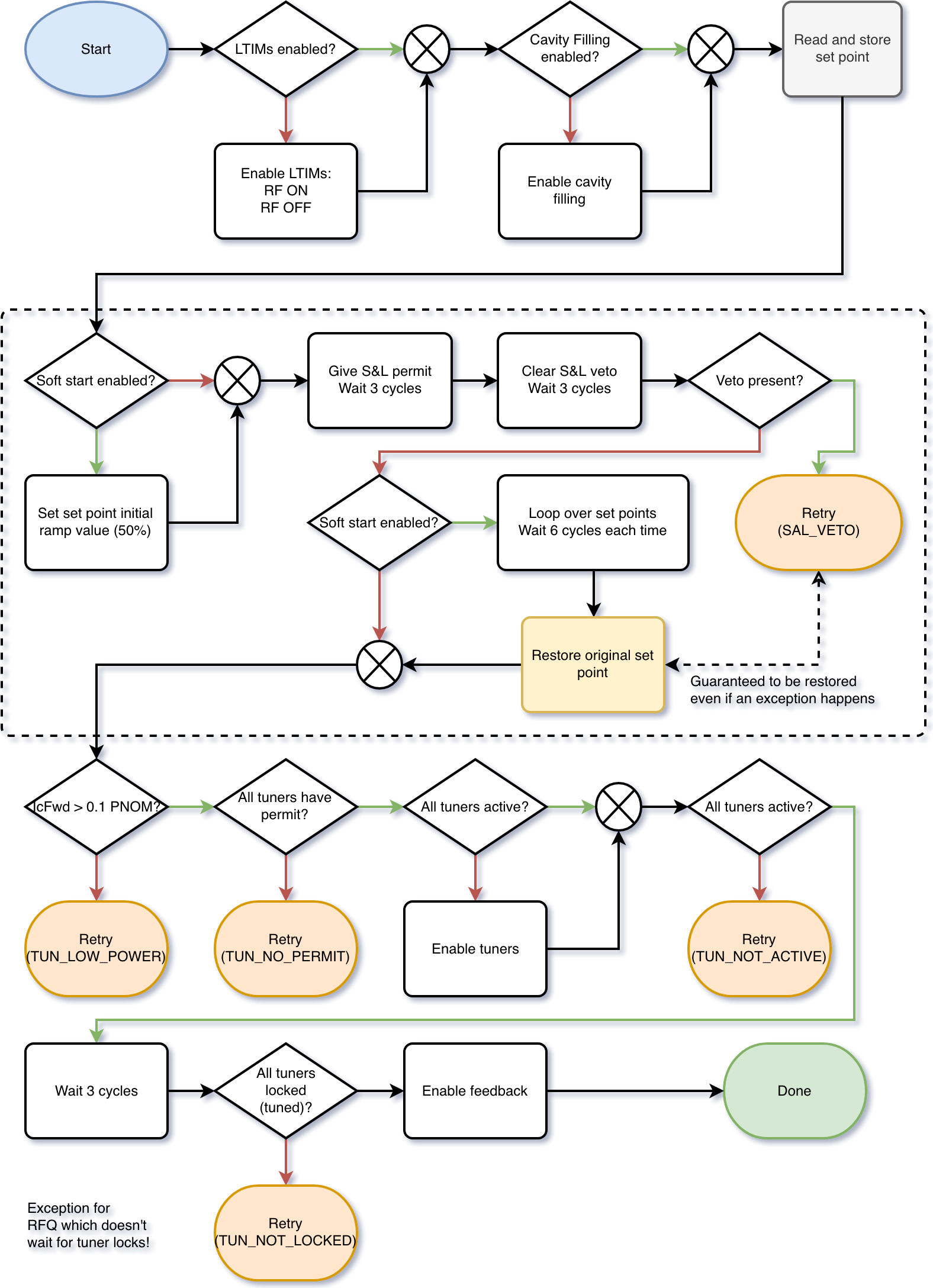}
    \caption{Flow chart of "LLRF on" sequence for Linac4.}
    \label{fig:sequencer-llrfon}
\end{figure}

The sequence of actions is complex and depends on the RF line characteristics. Use of Python, a real, object oriented programming language makes it possible to hide these complexities behind generic interfaces and makes development of the tool much easier. Figure \ref{fig:sequencer-llrfon} shows a flow graph of actions that turn the LLRF on. Together with the sequencer other tools have been created for setting up the LLRF, two of these have already been mentioned for the setting up of the AFF and Kalman predictor.

\section{CONCLUSION}

Linac4 LLRF system, both hardware and software, has reached a stable state that has allowed it to accelerate beams to the nominal 160 MeV energy with the simple PI controller in 2018. Our current efforts focus on improving performance and stability by fine-tuning the existing features, testing and introducing new concepts (LQG, AFF).

The LQG/Kalman regulator has now been deployed on all cavities except the debuncher and is presently tested with beam. 

Adaptive Feed-Forward, tested so far in the laboratory test stand, will also be commissioned in the machine this year. Introduction of varying real-life beam induced disturbances will verify the concept and allow us to select an optimal set of parameters.

Longitudinal painting, important for injection into the PSB, will also be tested. The presence of AFF will allow the voltage in the cavity to closely follow changes in voltage set-points as requested by physicists. 

When all systems are finally tested it will be possible to create a top-level operational software optimising all parameters of the LLRF at the same time. For now Python based prototypes allow us to rapidly develop and test ideas. This software will become the blueprint for the future, more permanent software solutions.

\section{ACKNOWLEDGEMENTS}
The design of the Linac4 LLRF has started in 2009. During the past ten years of development several persons have participated to the project:
A.~Bhattacharyya made the first simulations of the LLRF. G.~Hagmann designed the hardware, J.~Noirjean and J.~Galindo developed the original firmware. B.~Kremel designed both hardware and firmware for the tuner. We also want to thank K.~Fong from TRIUMF for interesting advice on the Adaptive Feed-Forward.

\end{document}